# Absence of Scaling in the Integer Quantum Hall Effect

N. Q. Balaban, U. Meirav and I. Bar-Joseph

*Braun Center for Submicron Research, Department of Condensed Matter Physics*

*Weizmann Institute of Science, Rehovot 76100, Israel*

Abstract

We have studied the conductivity peak in the transition region between the two lowest integer Quantum Hall states using transmission measurements of edge magnetoplasmons. The width of the transition region is found to increase *linearly* with frequency but remains finite when extrapolated to zero frequency and temperature. Contrary to prevalent theoretical pictures, our data does not show the scaling characteristics of critical phenomena. These results suggest that a different mechanism governs the transition in our experiment.

In highly disordered samples, the integer Quantum Hall Effect (IQHE) shows broad regions in magnetic field, $B$, where the longitudinal conductivity, $\sigma_{xx}$, vanishes, while the Hall conductance, $\sigma_{xy}$, is quantized in units of $e^2/h$. In the transition regions, $\sigma_{xy}$ goes from one quantized plateau to the next while $\sigma_{xx}$ goes through a peaked value and decreases again. Theoretically, this behavior was explained using the scaling theory of localization [1, 2]. The peak conductance occurs when the chemical potential, $\mu$, crosses the extended state at energy $E_c$, in the middle of the Landau level (LL). Near this energy, the localization length $\xi$ is predicted to diverge as $\xi \propto |E - E_c|^{-\chi}$, leading to metallic behavior, with $\chi = 7/3$ being a universal exponent. The transition regions are described as successive QH-metal-QH phase transitions whose widths approach zero in infinite samples in the limit of zero temperature and frequency. The effect of finite temperature, $T$, or frequency, $f$, is to smear the quantum phase transition, which acquires a finite width, $\Delta B$. Hence, the width of the peaks between Quantum Hall (QH) states is predicted to follow a power law behavior $\Delta B \propto T^\kappa$ or $\Delta B \propto f^\kappa$, with an exponent related to the dynamical exponent $z$ via $\kappa = 1/z\chi$ [2].

This picture has been tested in a large number of experiments [1, 3] mainly through dc conductance measurements of the temperature dependence of $\Delta B$. Although some experiments show an impressive range of temperatures where this power law behavior holds, e.g. Ref. 4, some show limits where scaling seems to fail [5]. In recent measurements of the temperature dependence of the longitudinal resistivity of the QH-insulator transition the authors concluded that their data cannot be described by a scaling form [6].

Similar broadening of the QH peaks is expected when the conductance is probed with a microwave signal. However, due to the experimental difficulties associated with this technique, only one experiment so far has tackled the frequency dependence of the broadening and has shown agreement with the power law predicted by scaling theory [7]. In this work, we measure the width of the $\nu=2 \rightarrow \nu=1$ transition region, $\nu$ the filling factor, with a novel technique using the excitation and detection of edge magnetoplasmons. The width of the transition region was found to increase *linearly* with the frequency of the excitation and to remain finite in the limit of $T \rightarrow 0$ and $f \rightarrow 0$, in contrast with the prediction of scaling theory. We also performed simultaneous measurements of the temperature dependence of the dc conductivity showing similar departure from scaling theory. We believe that these observations merit a re-examination of other experiments previously analyzed within the scaling theory framework [7, 8].

Edge magnetoplasmons (EMP) transmission measurements are performed by sending a high frequency signal on one side of the sample and detecting its phase and amplitude on the other side. The transmission of edge magnetoplasmons can be expressed as function of the dc conductivity tensor, using Volkov and Mikhailov derivation of the dispersion relation $\omega(k)=\omega'+i\omega''$ [9]. The real part of the dispersion, $\omega'=2\pi f$, has been shown to be governed by $\sigma_{xy}(\omega) \approx \sigma_{xy}$ [10-12], with a logarithmic correction [13, 14]. The attenuation of the EMP, $\omega''$, depends on an effective scattering time $\tau$. For our low mobility sample, $\omega'\tau \ll 1$ and $\text{Re}\sigma_{xx}(\omega) \approx \sigma_{xx}$. In this regime, $\omega''$ is less understood but has been shown to closely follow $\sigma_{xx}$ in the transition between QH plateaus [10, 11], in qualitative agreement with the ellipsoid density profile model [15] that predicts $\omega'' \propto \sigma_{xx}$ [16]. Although the transmitted EMP signal does not present a direct measurement of $\sigma_{xx}$, the width of the transition regions versus frequency can be extracted, as it depends only on the functional form of $\sigma_{xx}$. The main advantage of EMP transmission over microwave absorption, as a technique for measuring the frequency dependence of the transition regions, is its higher sensitivity. The measurements can therefore be performed at low input power level (< 1nW) and on relatively small samples, similar to the standard Hall bars used for dc measurements.

The samples used are GaAs/AlGaAs heterojunctions with mobility $\mu=34\,000$ cm$^2$/V s and carrier concentration $n_s=1.4\times10^{11}$ cm$^{-2}$ at 1.5 K, which are patterned in rectangular mesas of $164\times64\mu$m$^2$. The coupling of high frequencies to the system is done via two semi-rigid coaxial lines mounted on the top-load rod of a dilution refrigerator. Thermal activation measurements at $\nu=4$ (not shown) are used to verify that the electronic temperature matched that of the lattice, down to the lowest temperature. The coupling of microwaves to the sample is done through 50 $\Omega$ tapered coplanar lines as described in [13]. The end of the inner conductor is capacitively coupled to the mesa edge and the effect of multiple EMP reflections is suppressed by patterning an additional large ohmic contact on the edge opposite to the transmission path (inset of Fig. 1(b)). The detected signal arises from the transmission of the edge excitation along the free boundary of length $L\cong220$ µm. Utilizing a vector network analyzer, after two stages of amplification, allows simultaneous measurement of both amplitude and phase of the transmitted signal and the removal of the stray coupling between the exciting and detecting electrodes [13]. For each excitation frequency we verify that the transmitted signal is proportional to the exciting one, thus ruling out heating effects. The transmitted signal amplitude, $T_r$, is given by $T_r=A\exp(-\sigma_{xx} \times g(B))$ with $A$ the coupling coefficient between the coplanar line and the sample and $g$ being a function of $B$ that contributes to a smooth envelope signal. By measuring the magnetic field dependence of the stray coupling, we verify that the weak capacitive coupling between the coplanar line and the sample is not affected by the magnetic field. The uncertainty in $\Delta B$ introduced by $g$ is evaluated to be ~5%. To facilitate a simultaneous measurement of the dc conductance, a 100 µm wide Hall bar was patterned on the same chip. The Hall and longitudinal resistances are measured using standard lock-in technique with a 0.5 nA excitation current. The width of the peaks was unaffected by varying the measuring current from 0.1 to 1 nA.

A typical measurement of the raw transmitted signal, $T_r$, versus magnetic field is shown in the inset of Fig. 1(a) and the resulting attenuation $\alpha$ in Fig. 1(a), with $\alpha$ defined as $\alpha=-\log(T_r/T_r^{max})$. The simultaneous measurement of the dc resistivity of an adjacent Hall bar is plotted in Fig. 1(b). The attenuation is small in the QH states, where the longitudinal resistance vanishes, and is very large at the transition regions. Below $\nu$~0.5, the EMP signal is completely damped (Inset of 1(a)). The value of $\Delta B$ for the $\nu=2\rightarrow\nu=1$ transition shown in the figure is the full width at half maximum (FWHM).

The width of the ν=2→ν=1 transition region, Δ$B$, is plotted versus $f$ in Fig. 2 (solid squares), at 0.15 K. The behavior is clearly linear, namely, $\Delta B = af + b$. Since it had been pointed out in Ref. 8 that different definitions of Δ$B$ lead to different dependencies, we also plot the dependence of the full width at *quarter* maximum (crosses), still observing a very similar linear behavior [17]. As we increase the temperature (inset of Fig. 2), the value of Δ$B$ extrapolated to zero frequency rises and follows the temperature dependence of the dc conductivity. At the lowest temperature Δ$B$ is dominated by the frequency, whereas at high temperature, it is mainly determined by the temperature.

In order to verify that the absence of scaling apparent in the linear frequency dependence of Δ$B$ is an inherent property of our samples, we also measure the temperature dependence of Δ$B$ via dc resistance measurements (Fig. 3). For comparison, both linear and power law fit are plotted on a log-log scale in the inset of Fig. 3 (dashed and solid line respectively). Clearly, the linear fit is better, similarly to the frequency dependence, and the width of the transition remains finite as $T→0$. Such a behavior of the dc resistivity peak has also been reported in Ref. 5 for Si MOSFET.

These observations suggest that the scaling behavior is dominated by another mechanism. These results represent a significant departure from current, widely accepted views of QH phase transitions. While a rigorous theory that could explain these findings has yet to be developed, we present here a phenomenological model similar to the interpretation of the non local conductivity developed in Ref. 18. Despite the simplifications, the model reproduce the main features of our observations, as well as those of Ref. 6.

The longitudinal resistance in the transitions between LLs has been shown in a number of works to be described mainly by the backscattering of the highest occupied LL, while the contribution of the lower levels adds a constant conductance of $e^2/h$ for each fully occupied LL [19-21]. The resistivity of the highest occupied LL can be directly measured in a 'four-terminal' configuration when $\nu<1$. It is characterized by a single parameter [19], which can be modeled as an effective barrier with transmission $t$, where $t$ varies between 0 and 1. Using Landauer-Buttiker formalism [22], the resistivity characterizing the highest occupied level and derived in [23] is

$$\rho_{xx} = \frac{h}{e^2}\frac{1-t}{t}, \tag{1}$$

while the Hall resistance, $\rho_{xy}$ will remain quantized and equal to $h/e^2$. As the magnetic field is scanned the highest occupied LL crosses the Fermi level and $t$ will vary from 1 to 0 in an abrupt manner. The resulting $\rho_{xx}$ will then vary from 0 to $\infty$. At finite temperature, the step function is smeared by the Fermi function and the effective conductance of the barrier becomes $t(T) = \int(-\frac{\partial f}{\partial \varepsilon})t(\varepsilon)$. Assuming a step function for $t(\varepsilon)$ leads to:

$$\rho_{xx} = \frac{h}{e^2}\exp(\frac{\Delta\varepsilon}{kT}) \tag{2}$$

with $\Delta\varepsilon=\mu-\varepsilon_0-eV$. Here $\varepsilon_0$ is the energy of the 1-D level and $eV$ the height of the barrier; close to the transition $\Delta\varepsilon \propto \Delta B$. In order to translate $\rho_{xx}$ to the resistivity measured at the $\nu=2\rightarrow\nu=1$ transition, $\rho_{xx}^{2\rightarrow1}$, the contribution of the fully occupied level is made by adding $e^2/h$ to the off diagonal conductivity, in a procedure described in Ref. 21 which together with Eq. (2) leads to $\rho_{xx}^{2\rightarrow1} = \dfrac{h}{e^2} \dfrac{\exp(\dfrac{\Delta\varepsilon}{kT})}{4+\exp(\dfrac{2\Delta\varepsilon}{kT})}$. The peak value of $\rho_{xx}^{2\rightarrow1}$ is $h/4e^2$ and the width of the peak is calculated by subtracting the two solutions of $\rho_{xx}^{2\rightarrow1}=const.$ [24]. The outcome of this admittedly simplified analogy to a 1-D conductor is: a) the width of the peak $\Delta B$ is *linear* in $T$; b) $\rho_{xx}$ of the highest occupied LL is exponential in $\Delta\varepsilon$, thus in $\Delta B$ [6]. Once the transmission of the effective barrier is not an ideal step function in energy but has a finite width at $T=0$, similarly to the case of a real 1-D constriction, the width of the transition region remains finite at $T=0$. Indeed, in Ref. 25, the barrier is taken as a saddle point potential where tunneling accounts for the non zero transmission at $T=0$. The temperature dependence of our dc data (see Fig. 3) is in good agreement with this qualitative picture. The width of the peak remains finite as $T\rightarrow0$ and has a linear slope at higher $T$.

The frequency dependence can similarly be understood, using the framework of Tien and Gordon [26] extended to 1D in [27]. The contribution of the microwave signal to transmission over a barrier is explained in terms of photon-assisted tunneling, which can be seen as creating a new set of electron states at energies shifted by *mhf* with respect to the original levels, where *m* is an integer. For frequencies comparable to the temperature, the electron states are not resolved and result in an additional smearing of $t(\varepsilon)$, whose width is roughly $mhf$. The width of the peaks in the longitudinal resistance will vary linearly with frequency, in agreement with the data of Fig. 2. In the low frequency limit, the data converges to the value obtained from the dc measurements.

The temperature and frequency dependencies are qualitatively similar, namely, the width of the transition is linear both in temperature and frequency. The slope of $\Delta B$ versus $hf$ is about a factor of 5 larger than the slope versus $kT$. It appears that in some sense, the microwave signal is more efficient than the temperature in the broadening of the transition region, which remains to be explained by a more quantitative picture.

In conclusion, we have measured the frequency and temperature dependence of the $\nu=2 \rightarrow \nu=1$ transition region in the QH regime by measuring the transmission of edge magnetoplasmons. Contrary to earlier conclusions, the measurements indicate the absence of scaling in frequency and temperature. The results are further supported by the temperature dependence of the dc longitudinal resistance measured simultaneously. A simple model is presented where the thermal and frequency smearing of the energy levels accounts for the main features of the transition.

We would like to acknowledge useful discussions with Y. Imry, Y. B. Levinson, D. Shahar, E. Shimshoni, A. Stern and A. Yacobi. We thank M. Heiblum, R. dePiciotto and Hadas Strikman for careful reading of this manuscript.

This work has been supported by the MINERVA fund and by the Israeli Science Foundation administered by the Israeli Academy of Science and Humanities.

**Figure captions**

Fig. 1. (a) The attenuation, $-\log(T_r/T_r^{max})$, of the transmitted microwave signal, $T_r$, at a frequency $f$=2.386 GHz versus magnetic field, measured at 150 mK. Inset: The raw transmitted signal $T_r$. (b) The longitudinal resistivity, $\rho_{xx}$, at $T$=150 mK measured at low frequency (22 Hz) on an adjacent Hall bar, with an excitation current of 0.5 nA. The density extracted from the EMP measurement is slightly higher than that deduced from Hall measurements, probably due to spatial inhomogeneity.  Inset: A schematic view of the samples used in the transmission measurements. The mesa is marked in gray and the metallic surfaces in black. The EMP signal propagates along the upper boundary. A large ohmic contact (dashed) is patterned on the lower boundary in order to absorb signals reflected from the receiving electrode.

Fig. 2. The width $\Delta B$ of the attenuation peak in the $\nu=2\rightarrow\nu=1$ transition versus microwave frequency, at 150 mK.  The width is taken at half maximum (solid squares), and quarter maximum (crosses), both leading to a similar linear dependence.  The solid line is a linear fit.  At low frequencies, the width $\Delta B$ extrapolates to a finite value, consistent with that obtained from the resistivity measurements. Inset: the FWHM data

plotted on a log-log scale at 150, 330 and 700 mK. The dashed lines are plotted as a guide to the eye.

Fig. 3. The width $\Delta B$ of the $\nu=2\rightarrow\nu=1$ transition region versus temperature, obtained from resistivity measurements of $\rho_{xx}$ on the adjacent Hall bar. The width decreases linearly with temperature and saturates a low temperatures. Inset: Comparison between the linear (dashed) and power law fit (solid) of the high $T$ range on a log-log scale.

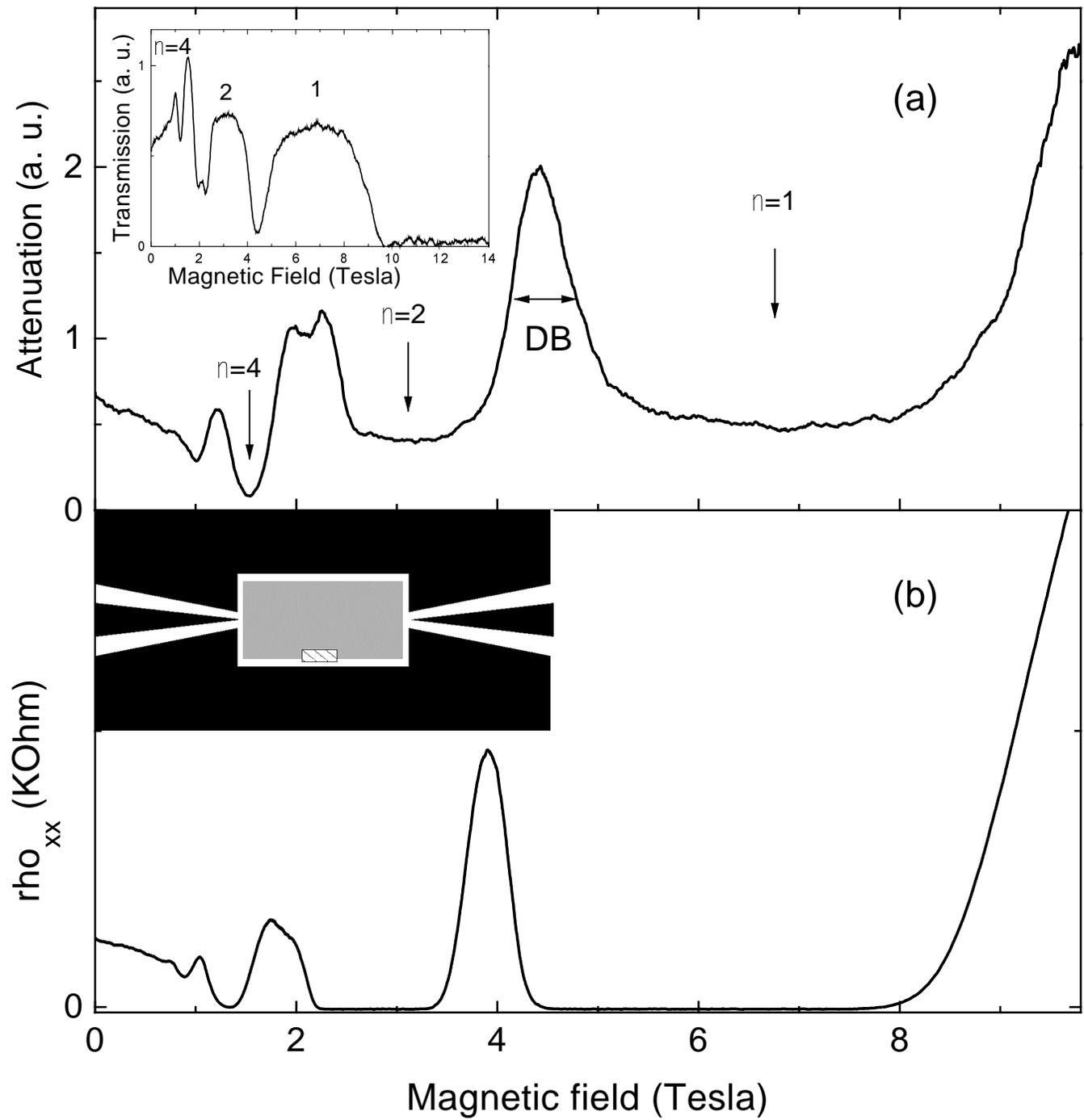

Fig. 1

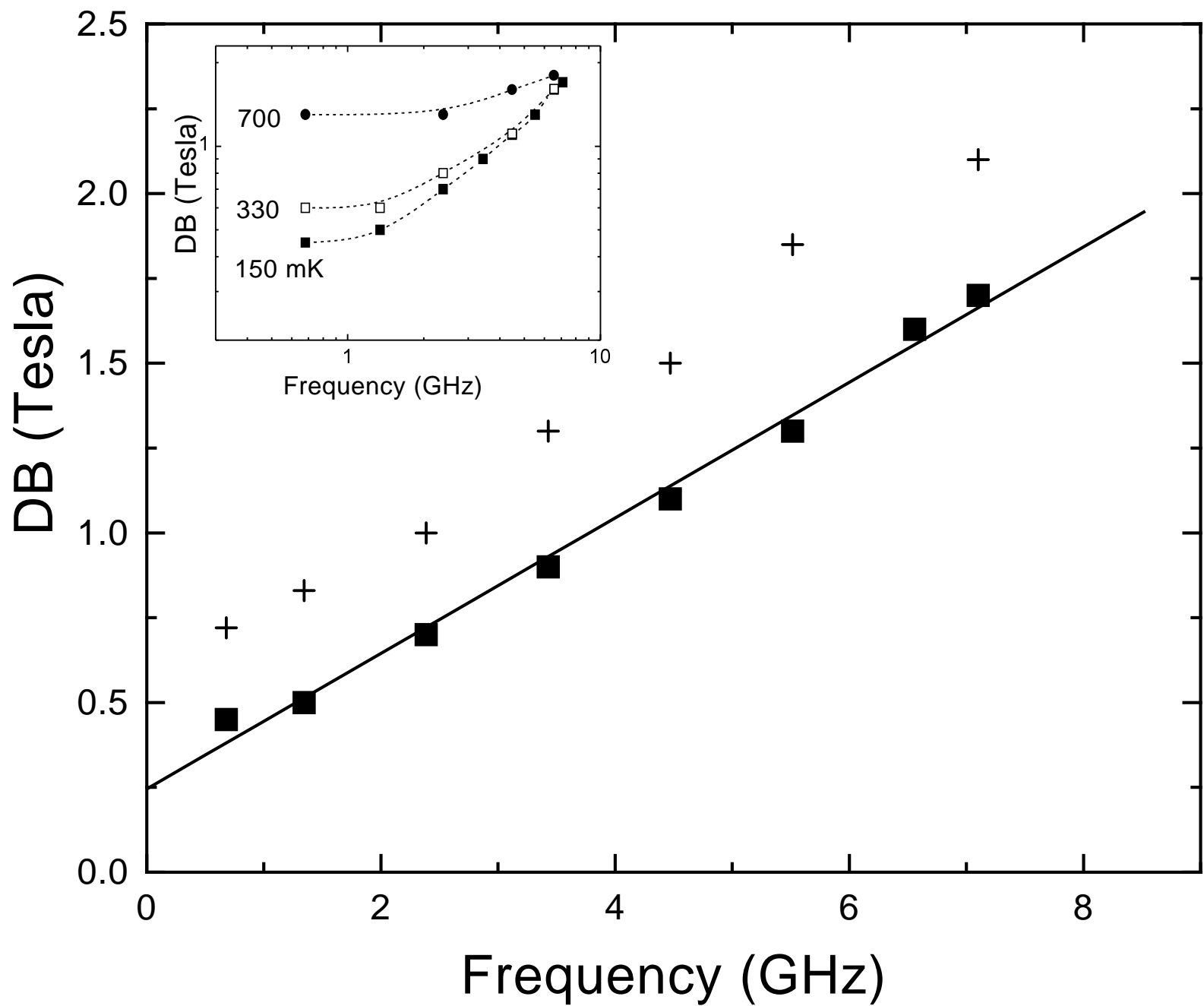

Fig. 2

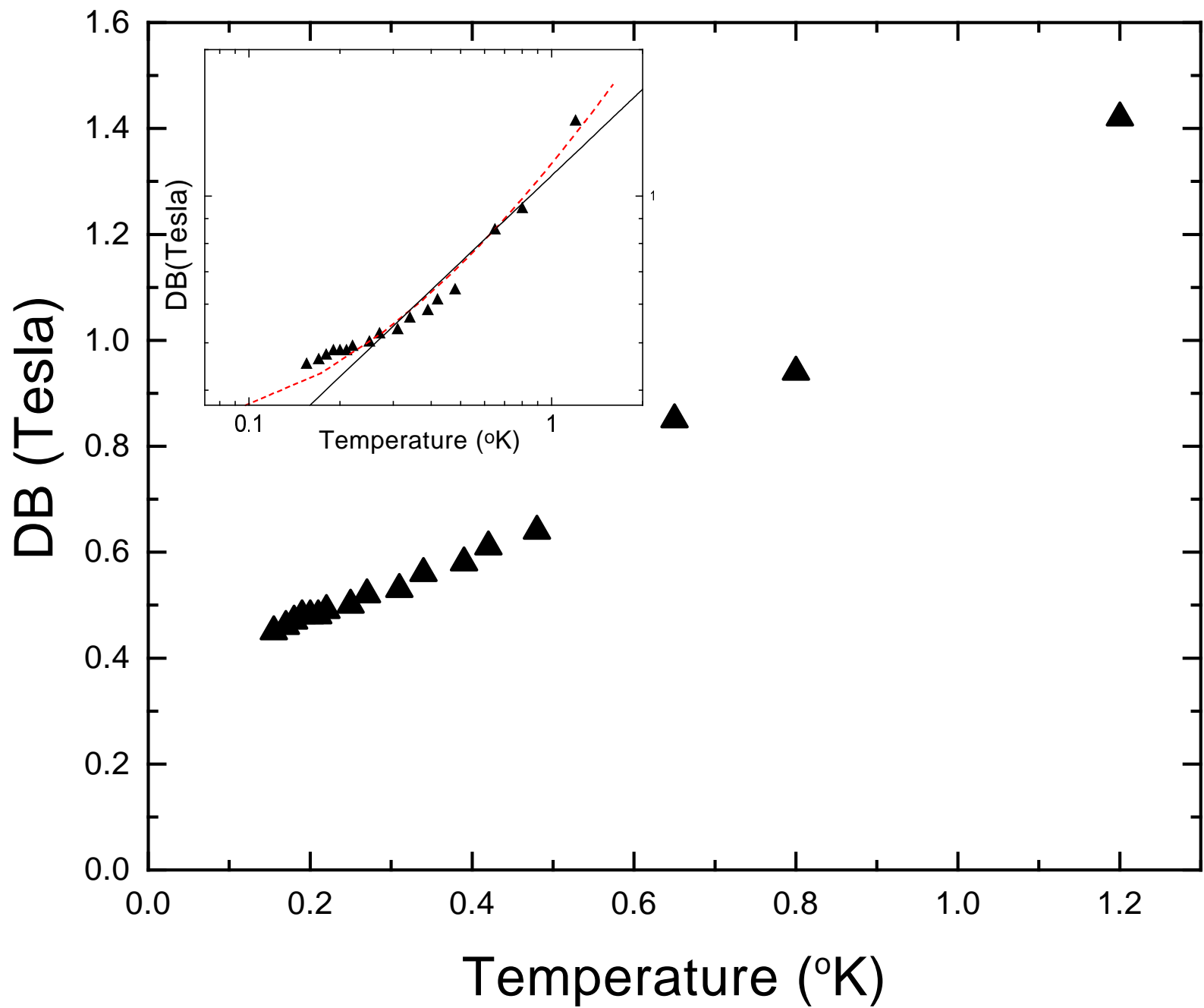

Fig. 3